\documentclass{PoS}

\usepackage{epsfig}

\PoS{PoS(LAT2005)349}

\title{The axial charge of the nucleon on the lattice and in chiral 
perturbation theory\thanks{Preprint DESY 05-175, Edinburgh 2005/08, LTH664}}

\ShortTitle{The axial charge of the nucleon on the lattice and in chiral 
 perturbation theory }

\author{Arifa Ali Khan$^a$,
        \speaker{Meinulf G\"ockeler}$^{a}$,
        Philipp H\"agler$^b$, 
        Thomas R. Hemmert$^b$, 
        Roger Horsley$^c$,
        Alan C. Irving$^d$,
        Dirk Pleiter$^e$,
        Paul E. L. Rakow$^d$, 
        Andreas Sch\"afer$^a$,
        Gerrit Schierholz$^{ef}$,
        Hinnerk St\"uben$^g$,
        Tim Wollenweber$^b$ and
        James M. Zanotti$^e$ 
        \\
        \llap{$^a$}  
        Institut f\"ur Theoretische Physik, Universit\"at Regensburg,
        93040 Regensburg, Germany \\
        \llap{$^b$}
        Physik-Department, Theoretische Physik, Technische Universit\"at
        M\"unchen, 85747 Garching, Germany \\
        \llap{$^c$}  
        School of Physics, University of Edinburgh, Edinburgh EH9 3JZ, UK \\
        \llap{$^d$}  
        Theoretical Physics Division, Department of Mathematical Sciences, 
        University of Liverpool, Liverpool L69 3BX, UK \\
        \llap{$^e$}  
        John von Neumann-Institut f\"ur Computing NIC, Deutsches 
        Elektronen-Synchrotron DESY, 15738 Zeuthen, Germany \\
        \llap{$^f$}  
        Deutsches Elektronen-Synchrotron DESY, 22603 Hamburg, Germany \\
        \llap{$^g$}  
        Konrad-Zuse-Zentrum f\"ur Informationstechnik Berlin, 
        14195 Berlin, Germany \\
        E-mail: 
        \email{meinulf.goeckeler@physik.uni-regensburg.de}} 
        \author{QCDSF-UKQCD Collaboration}

\abstract{We present recent Monte Carlo data for the axial charge of the 
nucleon obtained by the QCDSF-UKQCD collaboration for $N_f=2$ dynamical
quarks. We compare them with formulae from chiral perturbation theory in
finite and infinite volume and find a remarkably consistent picture.}

\FullConference{XXIIIrd International Symposium on Lattice Field Theory\\
		 25-30 July 2005\\
		 Trinity College, Dublin, Ireland}

\begin{document}

The QCDSF and UKQCD collaborations have generated ensembles 
of gauge field configurations using $N_f=2$ non-perturbatively $O(a)$ 
improved Wilson quarks and Wilson's plaquette action for the gauge fields. 
Here we want to discuss the results obtained for the axial
charge of the nucleon, $g_A$. Their interpretation is not straightforward
because the quark masses in the simulations are larger than in nature,
the volumes are somewhat smaller than infinity, the lattice spacings are
larger than 0 etc. So we need some guidance for the extrapolations 
towards the physical quark masses, the thermodynamic and continuum 
limits. Such guidance is provided by chiral effective field 
theory (ChEFT), which for selected quantities, e.g.\ for $g_A$, yields 
parameterisations of the dependence on the quark mass and the volume
which take into account the constraints imposed by (spontaneously
broken) chiral symmetry. The dependence on the lattice spacing $a$ 
can be included, but we shall not consider this possibility here. 
So we do not yet attempt to cope with the lattice artefacts remaining
even after $O(a)$ improvement.

If ChEFT can be successfully applied, we gain control over the chiral 
extrapolation and the approach to the thermodynamic limit. At the same 
time we can determine not only the physical value of the quantity of 
interest, $g_A$ in our case, but also some effective coupling constants. These
may occur in the ChEFT expressions for other observables and be of
phenomenological interest there. Establishing the link between Monte Carlo
results and ChEFT will thus enable us to extract considerably more
information from our simulations than just the physical value of the 
quantity under study. 

In its standard form, ChEFT describes low-energy QCD by means of an 
effective field theory based on effective pion, nucleon, {\ldots} fields.
Since the effective Lagrangian does not depend on the volume, 
besides the quark-mass dependence the very same Lagrangian governs 
also the volume 
dependence, and finite size effects can be calculated by evaluating the 
theory in a finite (spatial) volume. Thus the finite volume does not 
introduce any new parameters and the study of the finite size effects 
yields an additional handle on the coupling constants of ChEFT. 
The effective description will break down if the box length $L$ becomes 
too small, just as it fails for pion masses that are too large. 

The simulation parameters are listed in Table~\ref{tab:param}.
Note that we have two groups of three ensembles each which differ only 
in the volume.

\begin{table}[htb]
\begin{center}

\begin{tabular}{rllll}
\hline
 \multicolumn{1}{c}{Coll.} & \multicolumn{1}{c}{$\beta $} 
& \multicolumn{1}{c}{$\kappa_{\mathrm {sea}}$} 
& \multicolumn{1}{c}{volume} \\
\hline
 QCDSF   & 5.20 & 0.1342  & $16^3 \times 32$ \\
 UKQCD   & 5.20 & 0.1350  & $16^3 \times 32$ \\
 UKQCD   & 5.20 & 0.1355  & $16^3 \times 32$ \\[0.5cm]
 QCDSF   & 5.25 & 0.1346  & $16^3 \times 32$ \\
 UKQCD   & 5.25 & 0.1352  & $16^3 \times 32$ \\
 QCDSF   & 5.25 & 0.13575 & $24^3 \times 48$ \\[0.5cm]
 QCDSF   & 5.40 & 0.1350  & $24^3 \times 48$ \\
 QCDSF   & 5.40 & 0.1356  & $24^3 \times 48$ \\
 QCDSF   & 5.40 & 0.1361  & $24^3 \times 48$ \\
\hline
\end{tabular}
\hspace{0.5cm}
\begin{tabular}{rllll}
\hline
 \multicolumn{1}{c}{Coll.} & \multicolumn{1}{c}{$\beta $} 
& \multicolumn{1}{c}{$\kappa_{\mathrm {sea}}$} 
& \multicolumn{1}{c}{volume} \\
\hline
 UKQCD   & 5.29 & 0.1340  & $16^3 \times 32$ \\
 QCDSF   & 5.29 & 0.1350  & $16^3 \times 32$ \\[0.5cm]
 QCDSF   & 5.29 & 0.1355  & $12^3 \times 32$ \\
 QCDSF   & 5.29 & 0.1355  & $16^3 \times 32$ \\
 QCDSF   & 5.29 & 0.1355  & $24^3 \times 48$ \\[0.5cm]
 QCDSF   & 5.29 & 0.1359  & $12^3 \times 32$ \\
 QCDSF   & 5.29 & 0.1359  & $16^3 \times 32$ \\
 QCDSF   & 5.29 & 0.1359  & $24^3 \times 48$ \\
\hline
\rule{0pt}{5pt} & {} & {} & {} & {} \\
\end{tabular}

\end{center}
\caption{Simulation parameters.}
\label{tab:param}
\end{table}

We compute $g_A$ from forward proton matrix elements of the flavour-nonsinglet
axial vector current $ A_\mu^{u-d} =  \bar{u} \gamma_\mu \gamma_5 u 
- \bar{d} \gamma_\mu \gamma_5 d $:
\begin{equation}
\langle p,s | A_\mu^{u-d} | p,s \rangle = 2 g_A s_\mu \,.
\end{equation}
The required bare matrix elements are extracted from ratios of 3-point 
functions over 2-point functions in the standard fashion. Compared to
the computation of hadron masses, additional difficulties arise in the
calculation of nucleon matrix elements such as $g_A$: In general there
are quark-line disconnected contributions, which are hard to evaluate, the
operators must be improved and renormalised etc. Fortunately, in the limit
of exact isospin invariance, which is taken in our simulations, all
disconnected contributions cancel in $g_A$, because it is a 
flavour-nonsinglet quantity. The improved axial vector current is given by
\begin{equation}
A_\mu^{\mathrm {imp}}(x) =  \bar{q}(x) \gamma_\mu \gamma_5 q(x)
+ a c_A \partial_\mu  \bar{q}(x) \gamma_5 q(x) \,,
\end{equation}
and hence the improvement term, i.e.\ the term proportional to $c_A$, 
does not contribute in forward matrix elements. The renormalised improved
axial vector current can be written as
\begin{equation}
A_\mu = Z_A \left( 1 + b_A a m \right) A_\mu^{\mathrm {imp}}
\end{equation}
with the bare quark mass 
$m = \left( 1/\kappa_{\mathrm {sea}}- 1/\kappa_c \right)/(2a) $.

While the coefficient $b_A$ will be computed in tadpole improved one-loop
perturbation theory, we calculate the renormalisation factor $Z_A$ 
non-perturbatively by means of the Rome-Southampton method~\cite{rimom,reno}. 
Thus $Z_A$ is first obtained in the so-called 
RI'-MOM scheme. Using continuum perturbation theory we 
switch to the $\overline{\mbox{MS}}$ scheme. For sufficiently large
renormalisation scales $\mu$, $Z_A$ should then be independent of $\mu$.
However, unless $\mu \ll 1/a$ lattice artefacts may spoil this behaviour.
Since our scales do not always satisfy this criterion, we try to correct 
for this mismatch by subtracting the lattice artefacts perturbatively with
the help of boosted one-loop lattice perturbation theory. 
Some lattice artefacts still remain, but we can nevertheless estimate $Z_A$. 
In Table~\ref{tab:za} we compare our 
results with a recent determination of $Z_A$ by the ALPHA 
collaboration~\cite{alpha}.

\begin{table}[htb]
\begin{center}
\begin{tabular}{lllll}
\hline
$\beta$   &  5.20     & 5.25     & 5.29     & 5.40      \\
\hline
this work     &  0.765(5) & 0.769(4) & 0.772(4) & 0.783(4)  \\
ALPHA     &  0.719    & 0.734    & 0.745    & 0.767  \\
\hline
\end{tabular}
\end{center}
\caption{Values of $Z_A$ from this work and from the ALPHA collaboration.}
\label{tab:za}
\end{table}

\begin{figure}
\vspace*{-0.5cm}
\begin{center}
\epsfig{file=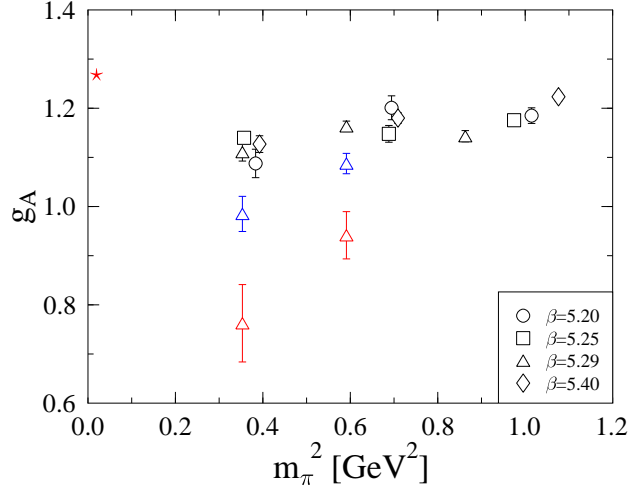,width=9.0cm}
\end{center}
\caption{Results for $g_A$. The red triangles correspond to a spatial box 
length $L \approx 1.0 \, \mbox{fm}$, the blue triangles belong to 
$L \approx 1.3 \, \mbox{fm}$, and for the black symbols the volumes 
are larger. The red star represents the physical point.} 
\label{fig:gadat}
\end{figure}

Our results for $g_A$ are plotted in Fig.~\ref{fig:gadat}. Here $m_\pi$
has been taken from the largest available lattice at each 
($\beta$, $\kappa_{\mathrm {sea}}$)
combination. The scale has been set by means of the force parameter $r_0$
with $r_0 = 0.467 \, \mbox{fm}$, and $r_0/a$ has been taken at the given 
quark mass. Obviously there are considerable finite size effects. A similar
volume dependence has already been observed in quenched 
simulations~\cite{sasaki}.

In order to describe or fit these data we use ChEFT. More specifically, 
we employ the so-called small scale expansion (SSE)~\cite{SSE}, 
which is one possibility to include explicit $\Delta$(1232) degrees of freedom
in ChEFT. The small expansion parameter in the SSE is called $\epsilon$,
and in $O(\epsilon^3)$ the mass dependence of $g_A$ is given 
for infinite volume by~\cite{hemmert}
\begin{equation} \begin{array}{l} \displaystyle
g_A^{SSE}(m_\pi^2) = g_A^0
+\left[4 C^{HB}(\lambda) -\frac{(g_A^0)^3}{16\pi^2F_\pi^2}
  -\frac{25c_A^2g_1}{324\pi^2F_\pi^2} +\frac{19c_A^2g_A^0}{108\pi^2F_\pi^2}
 \right]m_\pi^2
\\[0.5cm] \displaystyle {}
   -\frac{m_\pi^2}{4\pi^2 F_\pi^2}\left[(g^0_A)^3+\frac{1}{2}\,g^0_A\right]
 \ln{\frac{m_\pi}{\lambda }}
 +\frac{4c_A^2g_A^0}{27\pi\Delta_0F_\pi^2}\,m_\pi^3
\\[0.5cm] \displaystyle {}
  +\left[25c_A^2g_1\Delta_0^2-57c_A^2g_A^0\Delta_0^2
        -24c_A^2g_A^0m_\pi^2\right]
    \frac{\sqrt{m_\pi^2-\Delta_0^2}}{81\pi^2F_\pi^2\Delta_0}
     \arccos\frac{\Delta_0}{m_\pi}
\\[0.5cm] \displaystyle {}
 +\frac{25c_A^2g_1\left(2\Delta_0^2-m_\pi^2\right)}{162\pi^2F_\pi^2}
    \ln\frac{2\Delta_0}{m_\pi}
 +\frac{c_A^2g_A^0\left(3m_\pi^2-38\Delta_0^2\right)}{54\pi^2F_\pi^2}
  \ln\frac{2\Delta_0}{m_\pi}
   + O(\epsilon^4) \,,
\end{array}
\end{equation}
where $g_A^0$ denotes the chiral limit value of $g_A$.
This expression depends on several coupling constants, all referring to the 
chiral limit: $F_\pi$ is the pion decay constant with the physical value
of about 92.4 MeV, $\Delta_0$ denotes the real part of the 
$N \Delta$ mass splitting,
$c_A$ and $g_1$ are $N \Delta$ and $\Delta \Delta$ axial coupling constants,
respectively. Finally, $C^{HB}(\lambda)$ is a counterterm at the 
renormalisation scale $\lambda$, which can be expressed in terms of the
more conventional heavy-baryon couplings $B_9^r(\lambda)$ and 
$B_{20}^r(\lambda)$:
\begin{equation}
C^{HB}(\lambda) = B_9^r(\lambda) -2\,g_A^0 B_{20}^r(\lambda) \,.
\end{equation}
Evaluating the underlying ChEFT in a finite spatial volume yields an 
expression for the $L$ dependence of $g_A$~\cite{beane,tim}.

Phenomenology provides some information on the parameters appearing 
above. The analysis of (inelastic) $\pi$ $N$ scattering, in particular
the process $\pi N \to \pi \pi N$, suggests that choosing the physical
pion mass as the scale $\lambda$ one has~\cite{hemmert} 
\begin{equation}
B_9^r(\lambda = m_\pi^{\mathrm {phys}}) = (-1.4 \pm 1.2) \,\mbox{GeV}^{-2}
\quad , \quad B_{20}^r(\lambda = m_\pi^{\mathrm {phys}}) \equiv 0 \,.
\end{equation}
Therefore we set $\lambda = 0.14 \, \mbox{GeV}$ in the following and 
identify $C^{HB}(\lambda = m_\pi^{\mathrm {phys}}) =
 B_9^r(\lambda = m_\pi^{\mathrm {phys}})$. 
In the real world one has $\Delta_0 = 0.2711 \, \mbox{GeV}$, and from an 
$O(\epsilon^3)$ SSE analysis of the $\Delta$ width one finds $c_A = 1.5$.
At the physical pion mass we have $g_A = 1.267$, while 
$g_A^0 \approx 1.2$~\cite{hemmert}. Little is known about
$g_1$. In the SU(6) quark model one finds 
$g_1 = \frac{9}{5} g_A^0 \approx \frac{9}{5} 1.2 = 2.16 $. For
$F_\pi$ one expects in the chiral limit $F_\pi \approx 86.2 \, \mbox{MeV}$.

\begin{figure}
\vspace*{-0.5cm}
\begin{center}
\epsfig{file=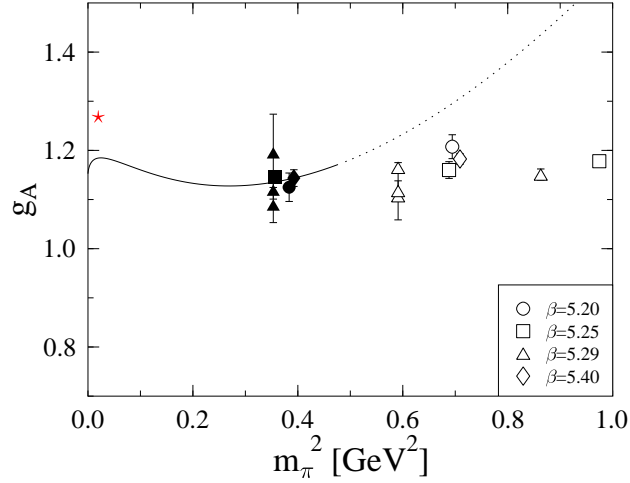,width=9.0cm}
\end{center}
\caption{Results for $g_A$ with the finite size correction subtracted. 
The curve represents the first fit described in the text. The data points 
shown as open symbols have not been fitted.} 
\label{fig:fita}
\end{figure}

Unfortunately, we cannot fit all parameters. So we fix 
$\Delta_0 = 0.2711 \, \mbox{GeV}$, $c_A = 1.5$,
$F_\pi = 86.2 \, \mbox{MeV}$ and fit $g_A^0$, $B_9^r$, $g_1$
taking into account only pion masses below approximately 600 MeV. 
In contrast to Ref.~\cite{hemmert} the physical point is not fitted.
We find $g_A^0 = 1.15(12)$, $B_9^r = - 0.71(18) \, \mbox{GeV}^{-2}$,
$g_1 = 2.6(8)$ with $\chi^2/\mathrm {dof} = 4.23/3$. Remarkably enough, 
these values are very well compatible with our phenomenological
prejudices above. In Fig.~\ref{fig:fita}
we plot the data with the finite size correction subtracted together with
the fit curve. So, if the fit would be perfect the data points which
differ only in the volume would collapse onto a single point.

\begin{figure}
\vspace*{-0.5cm}
\begin{center}
\epsfig{file=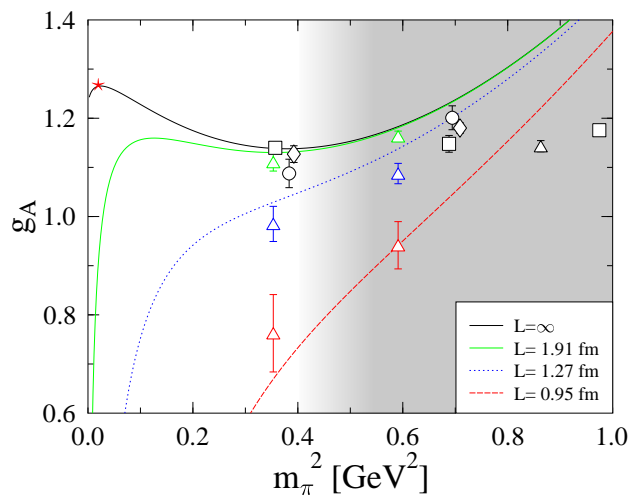,width=9.0cm}
\end{center}
\caption{Results for $g_A$ with curves for several values of $L$.}
\label{fig:fitb}
\end{figure}

The uncertainty in $g_A^0$ is large enough to cover the 
experimental point. To exemplify this circumstance we fix $g_A^0 = 1.225$,
a value well within the range favoured by the above fit, and use
only $B_9^r$ and $g_1$ as fit parameters. We find 
$B_9^r = - 0.66(17) \, \mbox{GeV}^{-2}$ and $g_1 = 3.0(3)$ with
$\chi^2/\mathrm {dof} = 4.61/4$. In Fig.~\ref{fig:fitb} we plot
the data without subtracting the finite size corrections. 
Using the results of the last fit we show curves not only for
$L=\infty$, but also for the values taken in the simulations for
$\beta=5.29$, $\kappa_{\mathrm {sea}} = 0.1359$.
Of course, many more variations of the fit procedure are possible, but
the overall pattern remains remarkably stable yielding $g_A^0 \sim 1.2$,
$B_9^r \sim - 0.5 \ldots - 0.7 \, \mbox{GeV}^{-2}$,  $g_1 \sim 3$ and
$F_\pi \sim 90 \, \mbox{MeV}$, in accordance with phenomenological
expectations.

\section*{Acknowledgements}

The numerical calculations have been performed on the Hitachi SR8000
at LRZ (Munich), on the Cray T3E at EPCC (Edinburgh)~\cite{allton},
and on the APEmille at NIC/DESY
(Zeuthen). This work is supported in part by the DFG (Forschergruppe
Gitter-Hadronen-Ph\"anomenologie) and by the EU
Integrated Infrastructure Initiative Hadron Physics under contract
number RII3-CT-2004-506078.

\end{document}